\documentclass[useAMS,usenatbib]{mn2e}
\usepackage{graphics,epsfig,lscape,txfonts,color}
\usepackage[english]{babel}
\newcommand{\mr}{\mathrm}

\newcommand{\hcm}[1]{$\times 10^{#1}$ cm$^{-2}$}

\def\ie{i.\,e.}                                      % i.e. (kursiv) \ie
\def\eg{e.\,g.}                                      % e.g. (kursiv) \eg

\def\xmm{\textit{XMM-Newton}}

\def\gx339{GX\,339--4}
\def\h1743{H\,1743--322}
\def\xte{XTE~J1650--500}

\title[Energy dependence of BLN break frequency in BHXRBs]{Energy dependence of the band-limited noise in black hole X-ray binaries\thanks{Based on observations obtained with \xmm, an ESA science mission with instruments and contributions directly funded by ESA Member States and NASA.}}
\author[H. Stiele, W. Yu]{H. Stiele \thanks{E-mail:
hstiele@mx.nthu.edu.tw} \thanks{Institute of Astronomy and Department of Physics, National Tsing Hua University, No.~101 Sect.~2 Kuang-Fu Road, Hsinchu, 30013, Taiwan}, W. Yu \\
Shanghai Astronomical Observatory and Center for Galaxy and Cosmology, 80 Nandan Road, Shanghai, 200030, China}% \\
\begin{document}

%\date{Accepted 1988 December 15. Received 1988 December 14; in original form 1988 October 11}
\date{2015 July 7}

\pagerange{\pageref{firstpage}--\pageref{lastpage}} \pubyear{2015}

\maketitle

\label{firstpage}

\begin{abstract}
Black hole low-mass X-ray binaries show a variety of variability features, which manifest themselves as narrow peak-like structures superposed on broad noise components in power density spectra in the hard X-ray emission. In this work we study variability properties of the band-limited noise component during the low-hard state for a sample of black hole X-ray binaries. We investigate the characteristic frequency and amplitude of the band-limited noise component and study covariance spectra. For observations that show a noise component with a characteristic frequency above one Hz in the hard energy band (4 -- 8 keV) we found this very same component at a lower frequency in the soft band (1 -- 2 keV).  This difference in characteristic frequency is an indication that while both the soft and the hard band photons contribute to the same band-limited noise component, which likely represents the modulation of the mass accretion rate, the soft photons origin actually further away from the black hole than the hard photons. Thus the soft photons are characterized by larger radii, lower frequencies and softer energies, and are probably associated with a smaller optical depth for Comptonisation up-scattering from the outer layer of the corona, or suggesting of a temperature gradient of the corona. We interpret this energy dependence within the picture of energy-dependent power density states as a hint that the contribution of  the up-scattered photons originating in the outskirts of the Comptonising corona to the overall emission in the soft band is becoming significant.
\end{abstract}


\section{Introduction}
Transient low-mass black hole X-ray binaries show distinct changes of their spectral and variability properties as they evolve during an outburst, that are interpreted as evidence for changes in the accretion flow and X-ray emitting regions. This behavior can be classified by different phenomenological states through which the black hole transient (BHT) evolves during an outburst \citep{2006ARA&A..44...49R,2010LNP...794...53B}. At the beginning of a typical outburst BHTs are usually in the low hard state (LHS) where their power density spectra (PDS) are dominated by band-limited noise (BLN) on which narrow quasi-periodic features (quasi-periodic oscillations; QPOs) are superimposed. During the evolution through the LHS and into the hard intermediate state (HIMS) the amplitude as well as the centroid frequency of the QPOs increases. In the energy range covered by the RXTE/PCA detector (\ie\ $>$2 keV) the PDS show a smooth transition from the LHS to the HIMS. This picture changes dramatically when softer energies are included, as they reveal the existence of two distinct power spectral states at soft and hard X-rays with the onset of the HIMS \citep{2013ApJ...770..135Y,2014MNRAS.441.1177S}. The transition to the soft intermediate state (SIMS) is marked by the appearance of another type of QPO, and an overall fractional rms in the 5 -- 10 per cent interval \citep{2005ApJ...629..403C,2011MNRAS.410..679M}. In the high soft state (HSS) the variability is even lower and the PDS is dominated by power-law noise. After experiencing a substantial decrease in luminosity during the HSS the BHT returns back to the LHS again usually passing the HIMS and SIMS \citep{2011MNRAS.418.1746S}.

In the LHS the X-ray emission is very noisy and the fractional rms can reach values up to 40--50 per cent.  The first detection of a difference in time scales between the soft and hard flux from a BHT was found in observations of the HIMS of GS\,1124--68 and\,GX 339--4 obtained with Ginga \citep{1997A&A...322..857B}.The shape of the PDS, which is dominated by BLN, can be fitted with a combination of Lorentzian components \citep{2002ApJ...572..392B}. Four main components have been identified (0.001 $<\nu<$ a few 100 Hz): a low-frequency one fitting the flat-top part, a peaked (sometimes QPO-like) component, and two broad Lorentzians at higher frequencies. These components are characterized by their characteristic frequency and by the amount of variability they contribute, where the latter one is measured as factional rms. Studying the fractional rms as a function of energy (the ``rms spectrum'') it was shown that it is either flat or decreases by a few per cent in the energy range covered by RXTE/PCA \citep[][and references therein]{2011BASI...39..409B}. This shape has been interpreted in the framework of Comptonization models \citep{2005MNRAS.363.1349G}. Using covariance spectra and ratios obtained from \xmm\ data, an increased disc blackbody variability with respect to the Comptonized emission was detected below 1 keV at time scales longer than one second, while on shorter time scales the disc variability is driven by variability of the Comptonized component, consistent with propagating models modified by disc heating at short time scales \citep{2009MNRAS.397..666W,2012MNRAS.427.2985C}. In addition to the BLN component type-C QPOs \citep{1999ApJ...526L..33W,2011MNRAS.418.2292M} can be present. They can be observed over a relatively large range of frequencies -- roughly from 0.01 to 30 Hz -- and their fractional rms is of the oder of 3 -- 15 per cent.

In recent years much work has gone into developing and exploring models to describe QPOs and noise components. An excellent summary of these models can be found in the extensive recent review of \citet[][see also references therein]{2014SSRv..183...43B}. BLN can be explained within the ``propagating fluctuation'' model \citep{1997MNRAS.292..679L}, where variability is caused by variations in the mass accretion rate which propagate through the disc, so that shorter time-scale variations either from coronal flares or fluctuation in the mass accretion rate or similar processes in the inner parts of the disc are superimposed on longer time-scale variations from further out \citep[\eg][]{,2001MNRAS.323L..26U}. BLN is also seen in neutron star low mass X-ray binaries. Likely indicators of the orbital frequency in the innermost accretion flow such as the kHz QPO Frequency in neutron star low mass X-ray binaries are found to vary with the flux corresponding to the time scale of the BLN in a manner similar to that established for variability on longer time scales which is due to modulation in the mass accretion rate, providing an independent evidence that the BLN corresponds to the modulation in the mass accretion rate which causes changes in the inner edge or characteristic frequencies in the accretion flow \citep{2009ApJ...690.1363C}.  

\begin{figure*}
\resizebox{\hsize}{!}{\includegraphics[clip,angle=0]{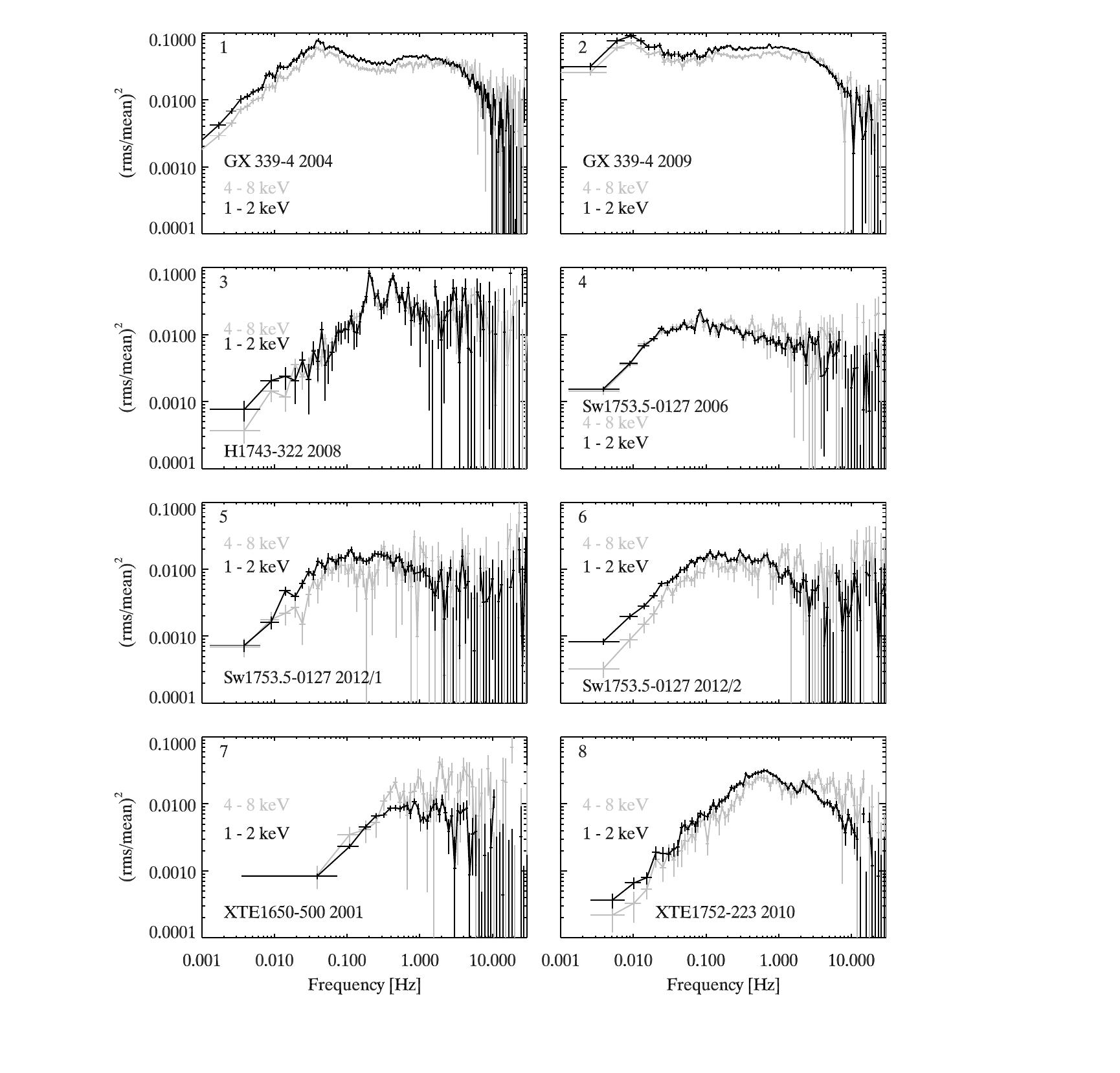}}
\caption{PDS of all eight observations in the soft (1 -- 2 keV) and hard (4 -- 8 keV) band. In observations where the characteristic frequency is at higher frequencies, the power in the hard band is systematically higher than the power in the soft band in the overall frequency range.} 
\label{Fig:PDS}
\end{figure*}

\section[]{Observations and data analysis}
\label{Sec:obs}
In this paper, we study the variability properties of the characteristic frequency and variability amplitude of the BLN in different energy bands during the LHS for a sample of BHTs. We used observations of GX\,339--4, H\,1743--322, XTE\,J1650--500, XTE\,J1752--223, and Swift\,J1753.5--0127. Details on the individual observations are given in Table~\ref{Tab:Obs}. All observations were taken during the LHS. 
We filtered and extracted the pn event files, using standard SAS (version 13.0.0) tools, paying particular attention to extract the list of photons not randomized in time. For our study we selected the longest, continuous exposure available in each observation (see Table~\ref{Tab:Obs}). We used the SAS task \texttt{epatplot} to investigate whether the observations are affected by pile-up, and in the case of pile-up excluded the column(s) with the highest count rate until the selection results in an observed pattern distribution that follows the theoretical prediction quite nicely. We selected single and double events (PATTERN$<=$4) for our study.
 
\subsection{Timing analysis}
We extracted power density spectra (PDS) in two different energy bands, namely 1 -- 2 and 4 -- 8 keV. After verifying that the noise level at frequencies above 30~Hz is consistent with the one expected for Poissonian noise \citep{1995ApJ...449..930Z}, we subtracted the contribution due to Poissonian noise, normalised the PDS according to \citet{1983ApJ...272..256L} and converted to square fractional rms \citep{1990A&A...227L..33B}. The PDS were fitted with models composed of zero-centered Lorentzians for BLN components, and Lorentzians for QPOs. The resulting power spectra in the two energy bands thus contain only source variability and can be compared with each other.

In addition, we derived rms spectra for each component of the PDS using the amplitude (normalisation) of the Lorentzians. Furthermore, we derived covariance spectra following the approach described in \citet{2009MNRAS.397..666W}. To study variability on shorter time scales we used 0.1~s time bins measured in segments of 4~s. For longer time scales 2.7~s time bins in segments of 270~s were used. As reference band we used the energy range between 1 and 4 keV, taking care to exclude energies from the reference band that are in the channel of interest.

\subsection{Spectral analysis}
\label{SubSec:SpecAna}
We extracted energy spectra of all observations, using the RAWX values given in table~\ref{Tab:Obs} for the source spectra. To extract background spectra we used columns 3$\le$RAWX$\le$5. In the case of burst mode observations we used RAWY $<$ 140 following the procedure outlined in \citet{2006A&A...453..173K}. We payed special attention to generate ARF files of the pile-up corrected source region, following the steps laid down in the \xmm\ Users Guide. As it is known that energy spectra obtained form \xmm\ EPIC-pn fast-readout modes can be affected by gain shift due to Charge-transfer inefficiency (CTI), which leads to an apparent shift of the instrumental edges visible at low energies, we also generated spectra applying the SAS task \textsc{epfast} to the data. Apart form the observation of \h1743\ no significant changes between the energy spectra obtained with and without \textsc{epfast} are noticeable at low energies, and a residual feature at $\sim$2 keV, related to small shifts in energy gain at the Si-K and Au-M edges of the instrumental response, remains present in the data obtained with \textsc{epfast} \citep[][]{2011MNRAS.416..311K}. As there are no significant changes of the energy spectra at low energies and as \textsc{epfast} leads to an over-correction of the CTI at higher energies \citep[see \eg\ ][]{2012MNRAS.422.2510W,2014MNRAS.441.1177S}, we fitted the spectra obtained without \textsc{epfast} and limited our spectral study to the 1 -- 10 keV energy range. 

\begin{table*}
\caption{Details of \xmm\ observations}
\begin{center}
\begin{tabular}{llrlrrrr}
\hline\noalign{\smallskip}
 \multicolumn{1}{c}{\#} & \multicolumn{1}{c}{Source} & \multicolumn{1}{c}{Obs.~id.} & \multicolumn{1}{c}{Date}  & \multicolumn{1}{c}{Mode$^{\ddagger}$} &  \multicolumn{1}{c}{Net Exp. [ks]}  &  \multicolumn{1}{c}{Exp.$^{\dagger}$ [ks]}  &  \multicolumn{1}{c}{RAWX$^{*}$} \\
 \hline\noalign{\smallskip}
1 & GX\,339--4 & 0204730201 & 2004 March 16 & T & 135.067 & 93.20 & 31 -- 36 \& 40 -- 43\\
\noalign{\smallskip}
2 & GX\,339--4 & 0605610201& 2009 March 26 & T & 33.529 & 32.234 & 31 -- 37 \& 39 -- 45\\
\noalign{\smallskip}
3 & H\,1743--322 & 0554110201 & 2008 Sep. 29 & T & 22.148 & 13.5 & 32 -- 35 \& 39 -- 44\\
\noalign{\smallskip}
4 & Swift\,J1753.5--0127 & 0311590901 & 2006 March 24 & T & 41.948 & 40.643 & 31 -- 45\\
\noalign{\smallskip}
5 & Swift\,J1753.5--0127 & 0691740201 & 2012 Sep. 10 & T & 40.091 & 37.890 & 31 -- 36 \& 39 -- 45\\
\noalign{\smallskip}
6 & Swift\,J1753.5--0127 & 0694930501 & 2012 Oct. 08 & T & 30.938 & 28.740 & 31 -- 36 \& 39 -- 45\\
\noalign{\smallskip}
7 & XTE\,J1650--500 & 0136140301 & 2001 Sep. 13 & B & 24.947 & 23.201 & 30 -- 44\\
\noalign{\smallskip}
8 & XTE\,J1752--223 & 0653110101 & 2010 April 06 & T & 41.917 & 19.700 & 30 -- 46 \\
\hline\noalign{\smallskip} 
\end{tabular} 
\end{center}
Notes: \\
$^{\ddagger}$: T for timing mode, B for burst mode\\ 
$^{\dagger}$: exposure used in this study\\
$^{*}$: detector columns from which source photons have been selected 
\label{Tab:Obs}
\end{table*}

\begin{figure}
\resizebox{\hsize}{!}{\includegraphics[clip,angle=0]{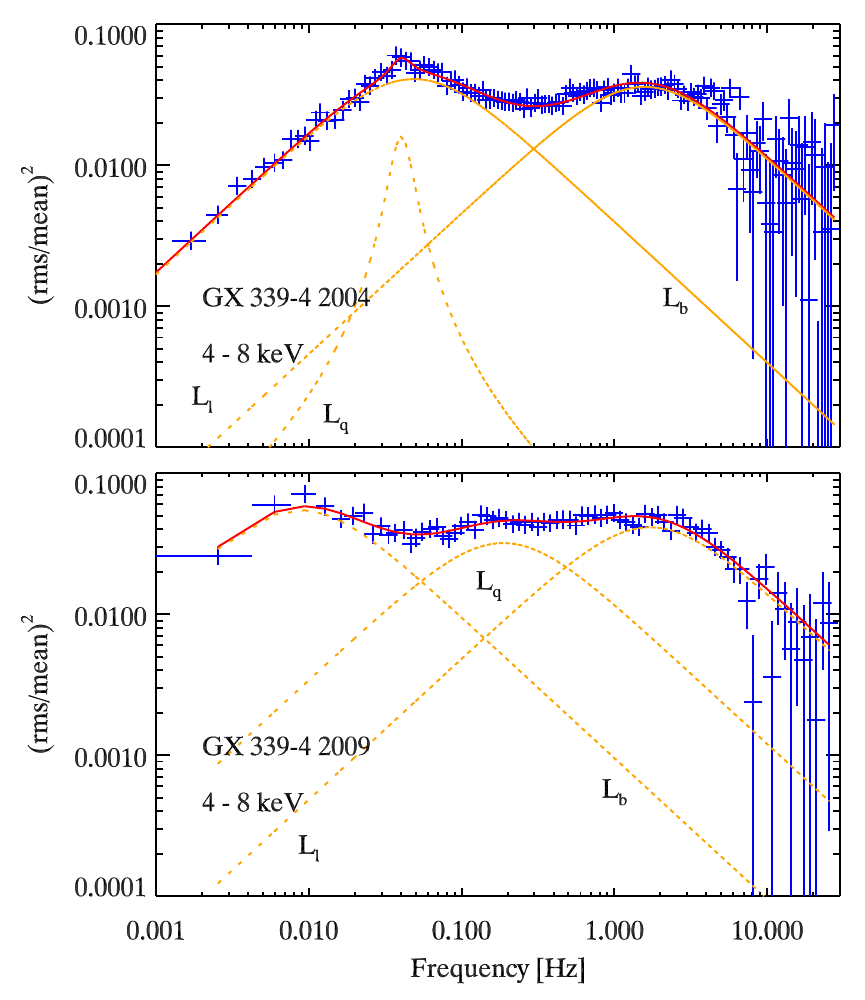}}
\caption{PDS of the 2004 (upper panel) and 2009 (lower panel) observations of GX\,339-4 in the hard (4 -- 8 keV) band to visualise the individual components (dashed orange lines). They are named following the nomenclature used in \citet{2002ApJ...572..392B} and \citet{2014SSRv..183...43B}. The component named L$_{\mr{q}}$ can be peaked (like in the upper panel) or broad (lower panel). For observations that only require two components to obtain decent fits the component L$_{\mr{q}}$ is not needed (which is the case for all observations of Swift\,J1753.5-0127; see Tab.~\ref{Tab:nub}).} 
\label{Fig:PDScomp}
\end{figure}

\begin{figure}
\resizebox{\hsize}{!}{\includegraphics[clip,angle=0]{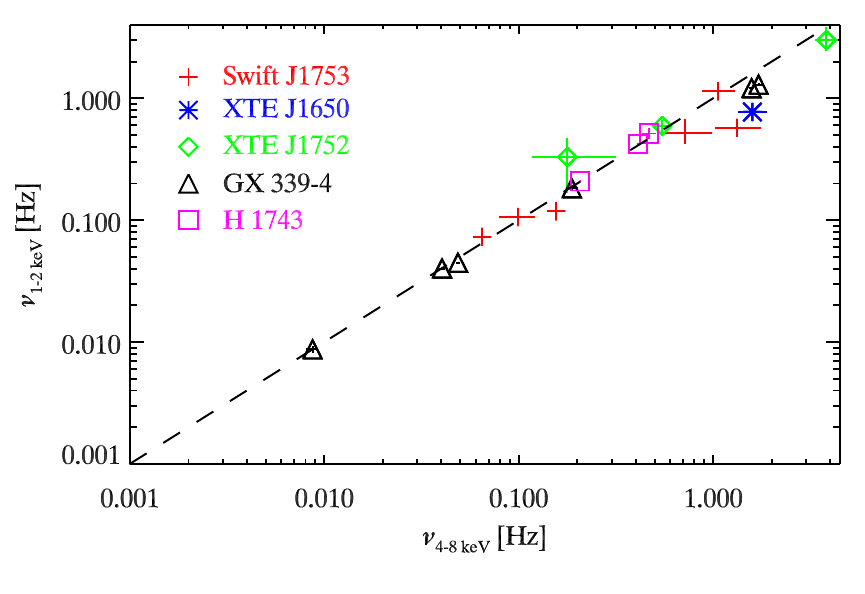}}
\resizebox{\hsize}{!}{\includegraphics[clip,angle=0]{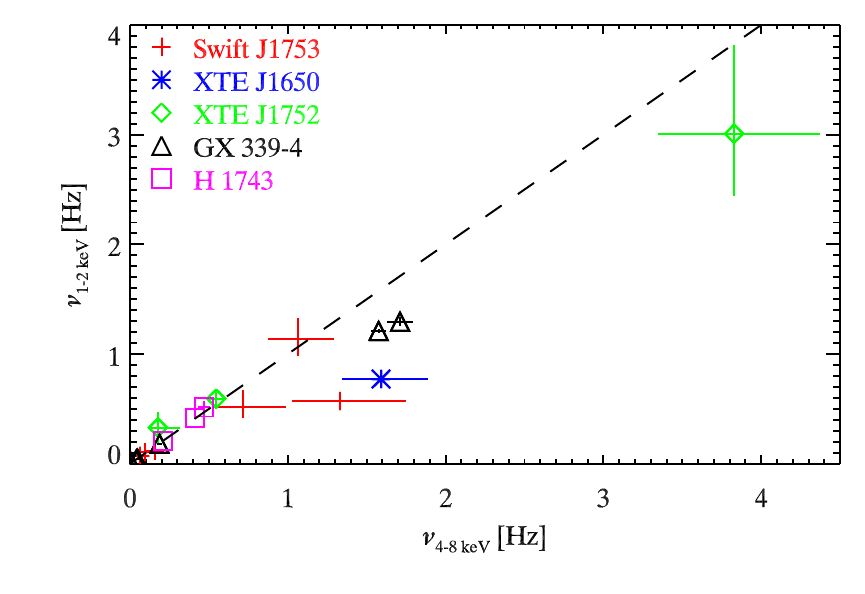}}
\caption{Characteristic frequency of the PDS components in the soft (1 -- 2 keV) versus hard (4 -- 8 keV) band, on double-logarithmic (upper panel) and linear scale (lower panel). The dashed line indicates equal frequencies. For almost all observations we find at least one data point which lies below the dashed line, \ie\ that we find one component for which the characteristic frequency obtained in the soft band is lower than the one obtained in the hard band. For the remaining noise components, the characteristic frequencies in both bands are equal within errors. Different symbols indicate individual sources.}
\label{Fig:numax}
\end{figure}

\begin{table}
\caption{List of characteristic frequencies for all observations in both energy bands.}
\begin{center}
\begin{tabular}{llrrl}
\hline\noalign{\smallskip}
\multicolumn{1}{c}{\#} & \multicolumn{1}{c}{Source} & \multicolumn{1}{c}{$\nu_{\mr{1 - 2 keV}}$} & \multicolumn{1}{c}{$\nu_{\mr{4 - 8 keV}}$} & L$^{\dagger}$\\
 \hline\noalign{\smallskip}
1 & GX\,339--4               &   $ 1.21\pm0.02 $ & $ 1.58\pm0.05 $ & L$_{\mr{l}}$\\
\noalign{\smallskip}
1 & GX\,339--4               &   $ 4.47\pm0.07 \times 10^{-2}$ & $ 4.87\pm0.10 \times 10^{-2} $ & L$_{\mr{b}}$\\
\noalign{\smallskip}
1 & GX\,339--4               &   $ 0.40\pm0.01 \times 10^{-1}$ & $ 0.40\pm0.02 \times 10^{-1}$ & L$_{\mr{q}}$\\
\noalign{\smallskip}
2 & GX\,339--4               &   $ 1.30\pm0.04 $ & $ 1.71\pm0.08 $ & L$_{\mr{l}}$\\
\noalign{\smallskip}
2 & GX\,339--4               &   $ 0.18\pm0.01 $ & $ 0.19\pm0.01 $ & L$_{\mr{q}}$\\
\noalign{\smallskip}
2 & GX\,339--4               &   $ 0.87\pm0.05 \times 10^{-2}$ & $ 0.87\pm0.06 \times 10^{-2} $ & L$_{\mr{b}}$\\
\noalign{\smallskip}
3 & H\,1743--322           &   $ 0.51_{-0.05}^{+0.06} $ & $ 0.49\pm0.04 $ & L$_{\mr{l}}$\\
\noalign{\smallskip}
3 & H\,1743--322           &   $ 4.20_{-0.08}^{+0.11}  \times 10^{-1}$ & $ 4.12_{-0.02}^{+0.04} \times 10^{-1}$ & L$_{\mr{qh}}$\\
\noalign{\smallskip}
3 & H\,1743--322           &   $ 2.08_{-0.03}^{+0.02} \times 10^{-1}$ & $ 2.07\pm0.02 \times 10^{-1}$ & L$_{\mr{q}}$\\
\noalign{\smallskip}
4 & Swift\,1753.5--0127 &   $ 1.14_{-0.16}^{+0.19} $ & $ 1.06_{-0.19}^{+0.23} $ & L$_{\mr{l}}$\\
\noalign{\smallskip}
4 & Swift\,1753.5--0127 &   $ 0.73\pm0.03\times 10^{-1} $ & $ 0.65\pm0.04\times 10^{-1} $ & L$_{\mr{b}}$\\
\noalign{\smallskip}
5 & Swift\,1753.5--0127 &   $ 0.52_{-0.10}^{+0.16} $ & $ 0.71_{-0.18}^{+0.27} $ & L$_{\mr{l}}$\\
\noalign{\smallskip}
5 & Swift\,1753.5--0127 &   $ 0.11\pm0.01 $ & $ 0.10\pm0.02 $ & L$_{\mr{b}}$\\ 
\noalign{\smallskip}
6 & Swift\,1753.5--0127 &   $ 0.57_{-0.06}^{+0.07} $ & $ 1.33_{-0.31}^{+0.42} $ & L$_{\mr{l}}$\\
\noalign{\smallskip}
6 & Swift\,1753.5--0127 &   $ 0.12\pm0.01 $ & $ 0.16\pm0.02 $ & L$_{\mr{b}}$\\
\noalign{\smallskip}
7 & XTE\,1650--500       &   $ 0.77\pm0.04 $ & $ 1.59_{-0.25}^{+0.30} $ & L$_{\mr{l}}$\\
\noalign{\smallskip}
8 & XTE\,J1752--223     &   $ 3.01_{-0.56}^{+0.81} $ & $ 3.83_{-0.48}^{+0.54} $ & L$_{\mr{l}}$\\
\noalign{\smallskip}
8 & XTE\,J1752--223     &   $ 0.59_{-0.09}^{+0.07} $ & $ 0.55\pm0.06 $ & L$_{\mr{q}}$\\
\noalign{\smallskip}
8 & XTE\,J1752--223     &   $ 0.33_{-0.16}^{+0.14} $ & $ 0.18_{-0.06}^{+0.14} $ & L$_{\mr{b}}$\\
\hline\noalign{\smallskip} 
\end{tabular} 
\end{center}
Notes: \\
L$^{\dagger}$: PDS component: L$_{\mr{l}}$, L$_{\mr{q}}$, and L$_{\mr{b}}$ are used in the way shown in Fig.~\ref{Fig:PDScomp}, while L$_{\mr{qh}}$ denotes the harmonic of the QPO\\
\label{Tab:nub}
\end{table}

\begin{figure}
\resizebox{\hsize}{!}{\includegraphics[clip,angle=0]{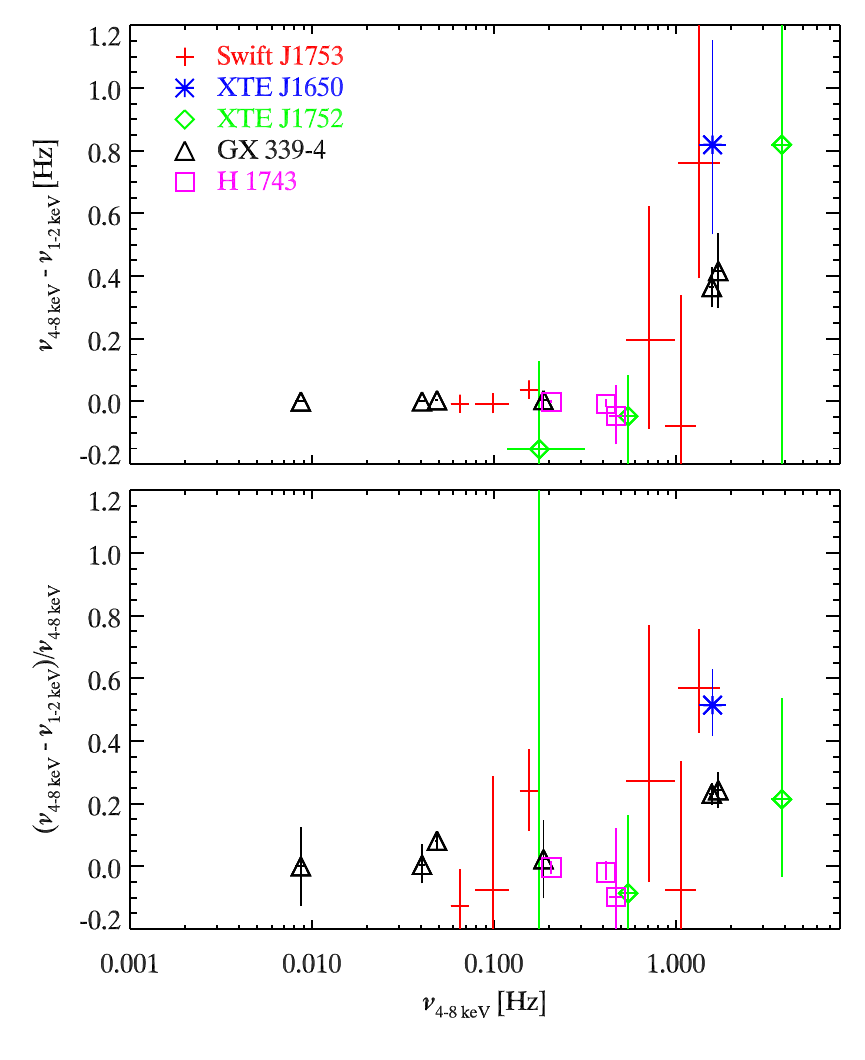}}
\caption{Absolute (upper panel) and relative (lower panel) change of the characteristic frequency between the hard and soft band versus the frequency in the hard band. Different symbols indicate individual sources.}
\label{Fig:diff}
\end{figure}

\section[]{Results}
\label{Sec:res}
\subsection{Studying timing properties}
\label{SubSec:time_prop}
Investigating the PDS of the individual sources in the two energy bands, we find that in general the overall shape of the components is consistent between the soft (1 -- 2 keV) and hard (4 -- 8 keV) band (see Fig.~\ref{Fig:PDS}). For some observations, an additional QPO or peaked noise component is present in one of the two energy bands but missing in the other one. In order to study the shapes of the PDS in a more quantitative way, we determined the characteristic frequency, defined by $\nu_{\mr{max}}=\sqrt{\nu^2+\Delta^2}$, where $\nu$ is the centroid frequency, and $\Delta$ is the half width at half maximum \citep{2002ApJ...572..392B}, of each component. In the case of BLN, where $\nu=0$, the characteristic frequency equals the break frequency. A visualization of the different components fitted to the hard band PDS of GX\,339--4 and named following the nomenclature used in \citet{2002ApJ...572..392B} and \citet{2014SSRv..183...43B} can be found in Fig.~\ref{Fig:PDScomp}. Figure~\ref{Fig:numax} (see also Tab.~\ref{Tab:nub}) shows the characteristic frequency obtained in the soft band versus the one derived from the hard band. Apart from the 2006 observation of Swift\,J1753.5-0127 and the observation of \h1743, we detect at least one component for which the characteristic frequency obtained in the soft band is lower than the one obtained in the hard band. For the remaining noise components, the characteristic frequencies in both bands are equal within errors. In cases where the characteristic frequency in the soft band is lower than in the hard band, this is always true for the component with the highest characteristic frequency (L$_{\mr{l}}$). \citet{2009MNRAS.397..666W} studied the variability of the 2004 observation of GX\,339--4 and the 2006 observation of Swift\,J1753.5--0127 in the 0.5 -- 1 keV and 2 -- 10 keV bands. They reported qualitatively, without obtaining characteristic frequencies through fits of the PDS, that in the case of GX\,339--4 the higher frequency component in the PDS appears ``to be shifted to even higher frequencies in the hard band'', while in the case of Swift\,J1753.5--0127 the ``soft and hard PDS appear to overlap more closely at higher frequencies". In a study of the SMBH system Ark 564 a similar energy dependence of the characteristic frequency of the higher frequency component has been found, while the lower frequency component did not show any energy dependence between the 0.6 -- 2 and 2 -- 10 keV bands \citep{2007MNRAS.382..985M}.  XTE J1753-223 that has been observed during outburst decay shows the highest break frequency and rather big error bars. A clear energy dependence of the characteristic frequency can be seen in the four observations that show break frequencies between 1 and 2 Hz in the hard band.

To rule out the possibility that the change of the characteristic frequency is mimicked by the emergence of a type-C QPO in the hard band, we added an additional Lorentzian to the fits of the hard band PDS and redetermined the characteristic frequencies. The allowed frequency range for the added Lorentzian was obtained from the relation of characteristic frequency and centroid QPO frequency presented in Fig.\ 11 of \citet{2002ApJ...572..392B}. For each observation, we used the lowest characteristic frequency found in the soft band to select the range of allowed centroid QPO frequencies. In addition the added Lorentzian had to be narrow. We found that the changes of the characteristic frequency caused by the added Lorentzian are within the errors of the values without added component. We estimated the significance of the added Lorentzian from its normalisation. This showed that the added Lorentzian is not significant at all (0.7 -- 1.4 $\sigma$).      

We also checked for the presence of additional disc variability at low frequencies in the soft band by adding a power law component to the model used to fit the PDS. A power law component was used to test for disc variability, as in the HSS, which is dominated by emission of the accretion disc, the PDS is consistent with power law noise at all frequencies \citep[see \eg][]{1994ApJ...435..398M,1997ApJ...474L..57C,2001MNRAS.321..759C,2006ARA&A..44...49R,2010LNP...794...53B}. In all observations fitting returned the normalisation of the power law component as zero, which implies that none of the PDS shows additional disc variability at low energies.

In Fig.~\ref{Fig:diff} we show the difference in the characteristic frequency between the hard and soft band, determined by $\delta\nu_{\mr{max}}=\nu_{\mr{max; 4-8 keV}}-\nu_{\mr{max; 1-2 keV}}$, versus the characteristic frequency in hard band. As can be seen from Fig.~\ref{Fig:diff} there is no clear correlation between the change of the characteristic frequency and the characteristic frequency itself. Using the characteristic frequency in the softer band instead of the one from the hard band on the x-axis leads to a shift to lower frequencies by $\delta\nu_{\mr{max}}$ for points with $\delta\nu_{\mr{max}}>0$ and to higher frequencies for those with $\delta\nu_{\mr{max}}<0$. In addition, Fig.~\ref{Fig:diff} also shows the relative changes of the characteristic frequency, defined as $\delta\nu/\nu_{\mr{max; 4-8 keV}}$ versus the characteristic frequency in hard band. Again no clear correlation between the parameters shown can be seen.

\begin{figure*}
\resizebox{\hsize}{!}{\includegraphics[clip,angle=0]{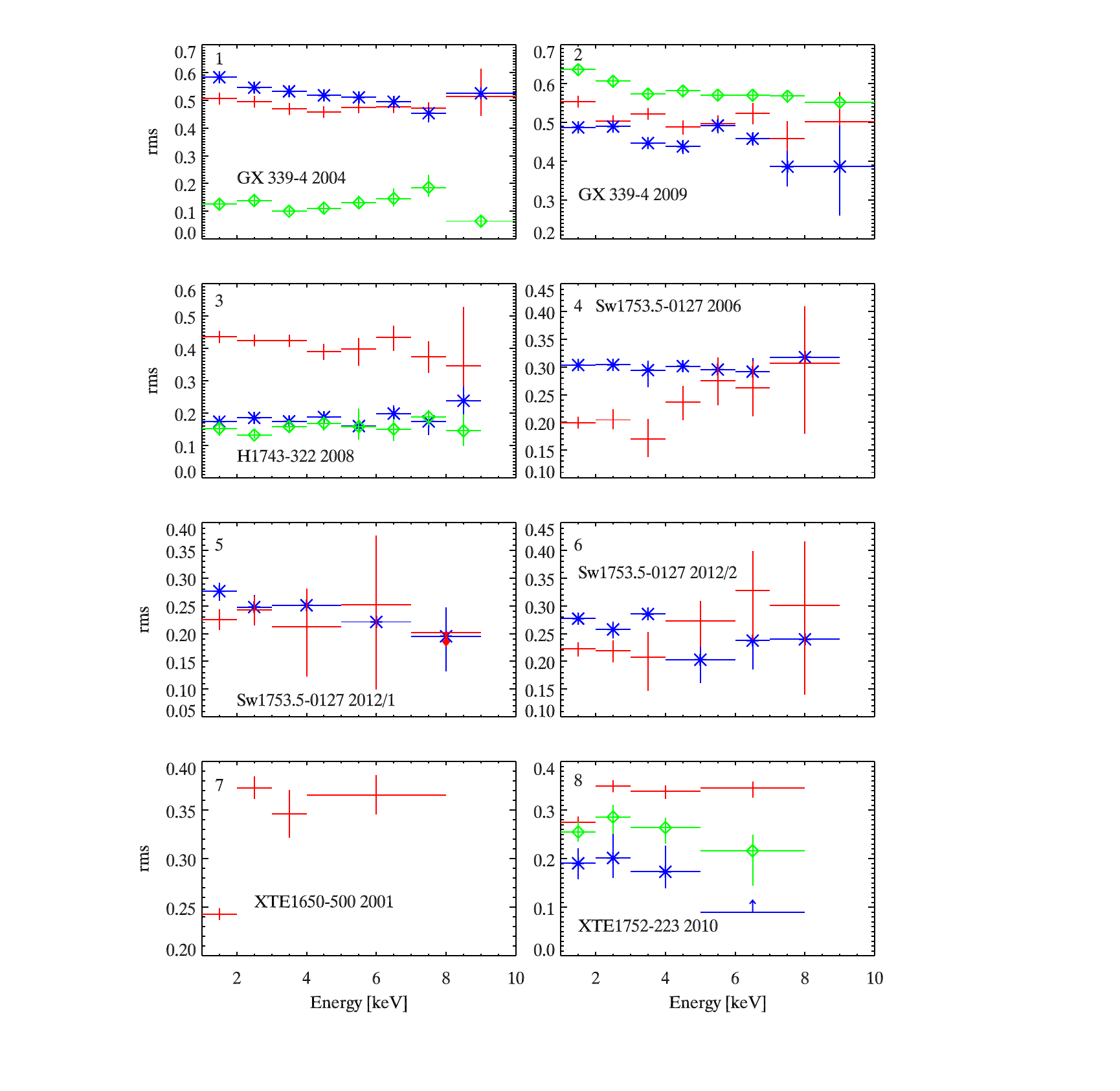}}
\caption{Rms spectra of all eight observations. Different symbols (colours) indicate different components of the PDS. The BLN component with the highest characteristic frequency (L$_{\mr{l}}$) is indicated by (red) crosses. The BLN component with the lowest characteristic frequency (L$_{\mr{b}}$) is indicated by (blue) stars, and the intermediate component (L$_{\mr{b}}$) is indicated by (green) diamonds, expect for \h1743\ where the QPO and its harmonic are indicated by (blue) stars and (green) diamonds, respectively.}
\label{Fig:rms_spec}
\end{figure*}

In addition to the energy dependence of the characteristic frequency, we investigate the energy dependence of the variability amplitude of the BLN. The rms spectra of all PDS components are shown in Fig.~\ref{Fig:rms_spec}. The BLN component with the highest characteristic frequency (L$_{\mr{l}}$) is indicated by (red) crosses. In the observations of \gx339, \h1743, and the September 2012 observation of Swift\,J1753.5--0127 this component does not show a strong energy dependence and it is either rather flat or slowly decreasing with increasing energy. This behaviour of the rms spectra has been observed in the hard state observations presented in \citet{2005MNRAS.363.1349G}. The remaining two observations of Swift\,J1753.5--0127 show a jump in variability around 4 -- 5 keV, while variability remains rather constant at lower and higher energies. A behaviour observed in the hard intermediate state \citep{2005MNRAS.363.1349G}. The observations of XTE\,J1650--500 and XTE\,J1752--223 show much lower variability in the 1 -- 2 keV band compared to the rather constant variability above 2 keV.

\subsection{Including energy spectra}
\label{Sect:espec}
We performed our investigation of the \xmm/EPIC-pn energy spectra within \textsc{isis} V.~1.6.2 \citep{2000ASPC..216..591H} in the 1 -- 10 keV energy range, taking into account a systematic error of 1\%. To obtain comparable spectral parameters, we fitted all spectra with the same model consisting of an absorbed disc blackbody plus thermal Comptonisation component \citep[\textsc{nthcomp};][]{1996MNRAS.283..193Z,1999MNRAS.309..561Z}, including a high-energy cut-off. The foreground absorption is modeled with \textsc{tbabs} \citep{2000ApJ...542..914W} and the values of  $n_{\mr{H}}$ are taken from the literature: $n_{\mr{H}}=7.8$\hcm{21} for XTE\,J1650--500 \citep{2002ApJ...570L..69M}, $n_{\mr{H}}=2.3$\hcm{21} for Swift\,J1753.5--0127 \citep{2013MNRAS.431.2341M}, $n_{\mr{H}}=5.0$\hcm{21} for \gx339\ \citep{2004MNRAS.351..791Z},  $n_{\mr{H}}=7.2$\hcm{21} for XTE\,J1752--223 \citep{2011arXiv1103.4312S} and $n_{\mr{H}}=1.6$\hcm{22} for H\,1743--322 \citep{2009MNRAS.398.1194C}. The temperature at the inner edge of the accretion disc is taken as temperature of the seed photons for the Comptonised component. In addition, a gaussian is added to model the residual visible in all observations at $\sim$2 keV (see Sect.~\ref{SubSec:SpecAna}).\@ In the case of \gx339\ a second gaussian is need to fit the Fe-emission line at 6.4 keV. In the case of \h1743\ we used the data with correction of CTI effects at low energies, and fitted those with the same model that has been used for \gx339. For \h1743\ the obtained fit implies a large disc normalization ($\sim5\times10^8$) and a very low inner edge temperature ($\sim10^{-3}$keV).
A more detailed study of the spectrum showed that the disc blackbody component is not need to obtain a decent fit. Hence, we removed this component, and fixed the seed temperature for the Comptonization model at the disc temperature mentioned above.  

\begin{figure}
\resizebox{\hsize}{!}{\includegraphics[clip,angle=0]{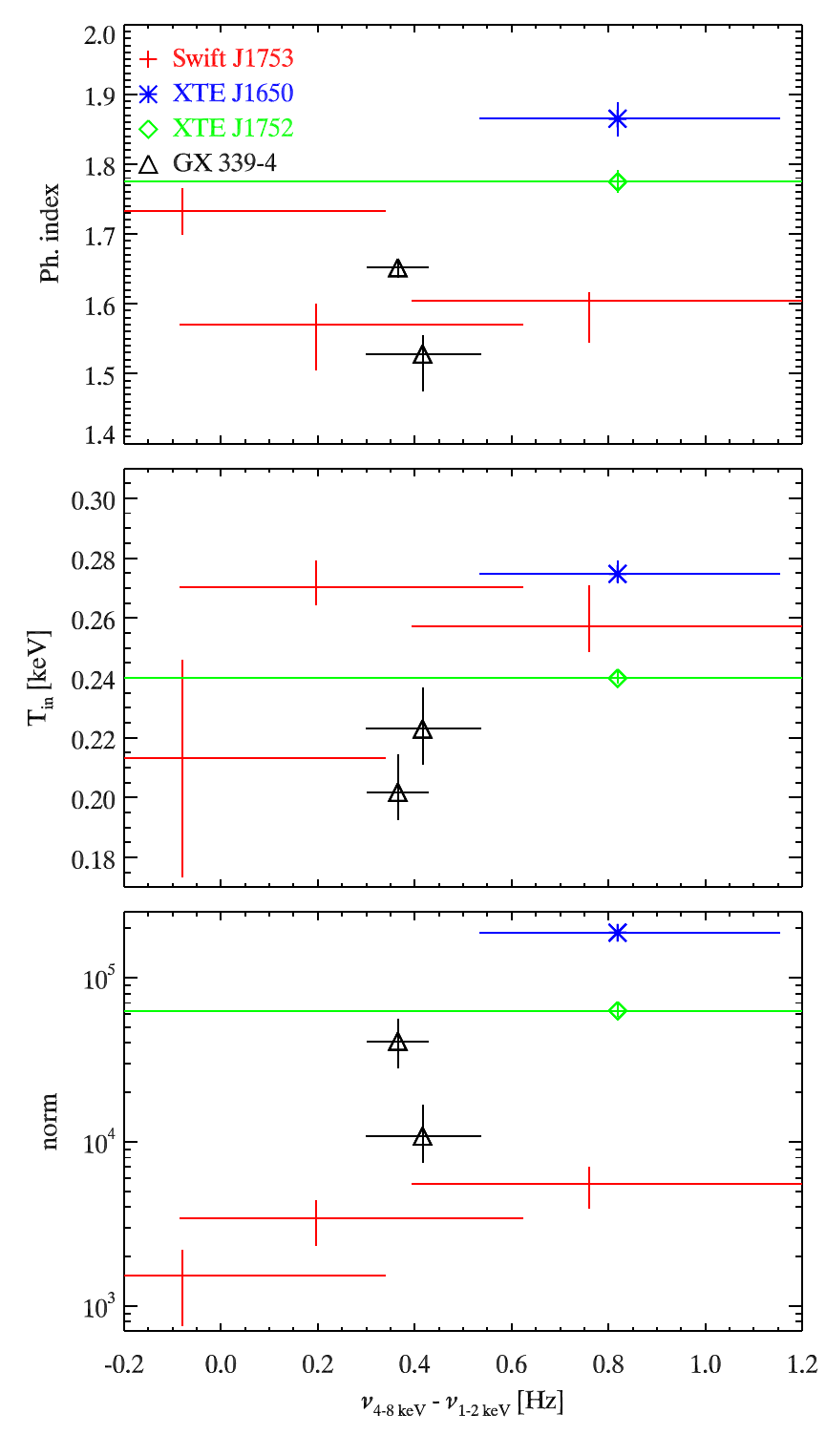}}
\caption{Behaviour of the photon index (upper panel), inner disc temperature (middle panel) and normalization of the disc component (lower panel) depending on the difference in the characteristic frequency $\delta\nu_{\mr{max}}$ of the component with the highest characteristic frequency. Different symbols indicate individual sources.}
\label{Fig:spec}
\end{figure}

\begin{figure}
\resizebox{\hsize}{!}{\includegraphics[clip,angle=0]{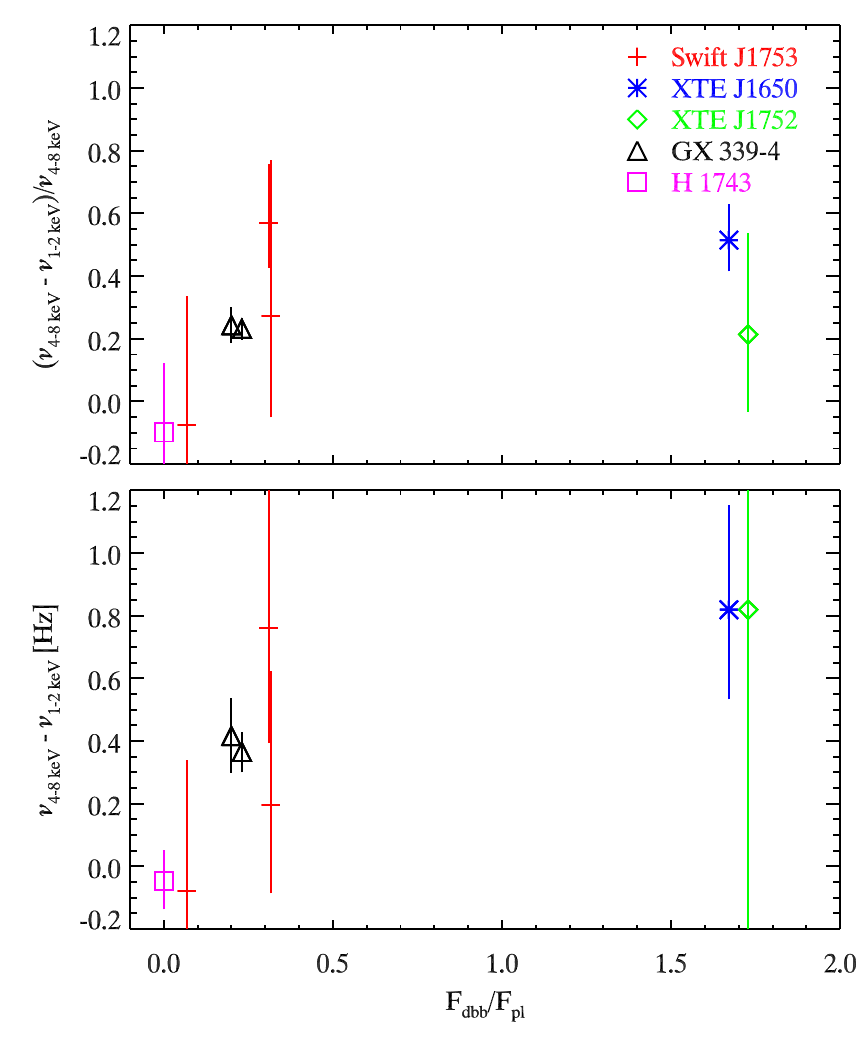}}
\caption{Absolute (upper panel) and relative (lower panel) change of the characteristic frequency between the hard and soft band of the component with the highest characteristic frequency versus the ratio of the flux contributed by the disc emission to the one of the Comptonized emission in the 1 -- 2 keV band. Different symbols indicate individual sources.}
\label{Fig:nuflux}
\end{figure}

Figure~\ref{Fig:spec} shows the behaviour of the photon index, inner disc temperature and normalization of the disc component depending on the difference in the characteristic frequency $\delta\nu_{\mr{max}}$ of the component with the highest characteristic frequency. The values for \h1743\ are omitted in this figure as no disc component was need to obtain an acceptable fit for this source. For the three observations of Swift\,J1753.5-0127 a slide increase in the disc normalization with increasing $\delta\nu_{\mr{max}}$ is visible. This behaviour is not observed in the two observations of \gx339, which span a smaller range in $\delta\nu_{\mr{max}}$ than the ones of Swift\,J1753.5-0127. Taken into account all observations again no clear correlation is evident.

Having a look at the ratio of the flux contributed by the disc blackbody component to the one coming for the Componized component in the soft (1 -- 2 keV) band (the values are given in the last row of Table~\ref{Tab:SpecPar}), we find that this ratio is below $\sim$7\% for the two observations for which the characteristic frequencies are consistent within errors (observation of \h1743\ and 2006 observation of Swift\,J1753.5$-$0127), while it is above 20\% for those observations where we found $\delta\nu_{\mr{max; 1-2 keV}}<\delta\nu_{\mr{max; 4-8 keV}}$. In the case of XTE\,J1650--500 and XTE\,J1752--223, the ratio exceeds one hundred per cent, which means that the flux of the disc component contributes more than 50\% of the total flux in the soft band. For  XTE\,J1752--223 this high value might be attributed to the fact that the source is observed during outburst decay, but this is not true for XTE\,J1650--500 which was observed during outburst rise. In Fig.\ref{Fig:nuflux} we show the flux ratio versus the absolute and relative  difference in the characteristic frequency of the component with the highest characteristic frequency. Even ignoring the two data points with a flux ratio larger than one, there no clear correlation that a larger absolute or relative change of the characteristic frequency would be related to a larger flux ratio. 

\begin{table*}
\caption{Spectral parameters}
\begin{center}
\begin{tabular}{lllllllll}
\hline\noalign{\smallskip}
 \multicolumn{1}{c}{param.} & \multicolumn{1}{c}{GX\,339/04} & \multicolumn{1}{c}{GX\,339/09}  & \multicolumn{1}{c}{Sw1753/06} &  \multicolumn{1}{c}{Sw1753/12/1}  &  \multicolumn{1}{c}{Sw1753/12/2}  &  \multicolumn{1}{c}{XTE1650}  &  \multicolumn{1}{c}{XTE1752} &  \multicolumn{1}{c}{H\,1743} \\
 \hline\noalign{\smallskip}
N$_{\mr{dbb}}$ & $40922_{-13085}^{+15215}$ & $10825_{-3448}^{+5996}$  & $1526_{-771}^{+676}$ & $3434_{-1102}^{+972}$ & $5526_{-1630}^{+1511}$ & $187432_{-14993}^{+17585}$ & $62815_{-5511}^{+6523}$ & -- \\
\noalign{\smallskip}
T$_{\mr{in}}$ [keV]& $0.202_{-0.009}^{+0.013}$ & $0.223_{-0.012}^{+0.014}$ & $0.213_{-0.040}^{+0.033}$ & $0.270_{-0.006}^{+0.010}$ & $0.257_{-0.009}^{+0.014}$ & $0.275_{-0.003}^{+0.004}$ & $0.234_{-0.002}^{+0.003}$ & 0.001$^{\natural}$\\
\noalign{\smallskip}
E$_{\mr{coff}}$ [keV]& $7.6\pm0.2$ & $7.4\pm0.2$ & $>9.3$ & $7.2\pm0.2$ & $7.3_{-0.2}^{+0.1}$ & $7.7\pm0.4$ & $6.8_{-0.1}^{+0.2}$ & $1.73_{-0.02}^{+0.04}$\\
\noalign{\smallskip}
E$_{\mr{fold}}$ [keV]& $17.8_{-3.8}^{+6.5}$ & $19.8_{-2.4}^{+5.6}$ & $-$ & $15.1_{-3.4}^{+3.9}$ & $15.4\pm2.2$ & $20.8_{-5.1}^{+7.4}$ & $35.5_{-4.6}^{+5.7}$ & $5.1_{-0.8}^{+1.3}$\\
\noalign{\smallskip}
N$_{\mr{nthc}}$ & $0.329_{-0.014}^{+0.009}$ & $0.199_{-0.010}^{+0.006}$ & $>0.06$ & $0.138_{-0.009}^{+0.006}$ & $0.167_{-0.008}^{+0.004}$ & $1.640_{-0.072}^{+0.070}$ & $0.226_{-0.005}^{+0.006}$ & $0.087_{-0.002}^{+0.003}$\\
\noalign{\smallskip}
$\Gamma$ & $1.65\pm0.01$ & $1.53_{-0.05}^{+0.03}$ & $1.73\pm0.03$ & $1.57_{-0.07}^{+0.03}$ & $1.60_{-0.06}^{+0.01}$ & $1.87_{-0.03}^{+0.02}$ & $1.77\pm0.02$ & $1.03_{-0.01}^{+0.03}$\\
\noalign{\smallskip}
kT$_{\mr{e}}$ [keV]& $4.6_{-0.7}^{+1.7}$ & $7.6_{-2.2}^{+7.6}$ & $>9.5$ & $8.0_{-3.0}^{+17.5}$ & $7.1_{-1.6}^{+11.4}$ & $>6.1$ & $>7.8$ & $14.1_{-1.4}^{+5.1}$\\
\noalign{\smallskip}
$\chi^2$/dof& $46.7/76$ & $48.3/76$ & $54.2/79$ & $92.6/79$ & $91.0/79$ & $83.7/79$ & $114.9/79$ & $75.0/78$\\
F$_{\mr{dbb}}$/F$_{\mr{pl}}^{\dagger} [\%]$ & 23.1 & 20.0 & 6.8 & 31.6 & 31.0 & 167.1 & 172.8 & 0\\
\hline\noalign{\smallskip} 
\end{tabular} 
\end{center}
Notes: \\
$^{\dagger}$: ratio of the flux contributed by the disc emission to the one of the Comptonized emission in the 1 -- 2 keV band\\
$^{\natural}$: value used as seed temperature in the \textsc{nthcomp} model (fixed) 
\label{Tab:SpecPar}
\end{table*}

\subsection{Covariance spectra}
To investigate variability of spectral components we derived covariance spectra on long (segments of 270~s with 2.7~s time bins) and short (segments of 4~s with 0.1~s time bins) time scales (Fig.~\ref{Fig:cove_spec}). The covariance spectra have been fitted with the same model used for the averaged energy spectra in Section~\ref{Sect:espec}. We adjusted the binning of the response files to the number of bins present in the covariance spectra using the ftool \textsc{rbnrmf}. In the cases where 50 bins have been used (\gx339\ and \h1743) the rebinding led to an artificial feature consisting of a spike and a trough-like structure at energies between 5.8 and 6.8 keV (the energy range of the Iron K$\alpha$ emission). Hence, this energy range has been ignored in the fits of the covariance spectra. For both observations of \gx339\ it is not possible to obtain formally acceptable fits as the data points stagger around the obtained best fit at energies below 2 -- 2.5 keV. In the spectral fitting of all covariance spectra a systematic uncertainty of one per cent has been included.

The covariance spectra of XTE\,J1650--500 and XTE\,J1752--223 do not require a high-energy cut-off component to obtain acceptable fits. For these two observations the photon indices and inner disc temperatures obtained from the covariance spectra are larger than those obtained from the mean energy spectra. 

For the remaining observations a comparison of the behaviour of the photon index and inner disc temperature between short and long term covariance spectra as well as between covariance and mean energy spectra can be found in Table~\ref{Tab:SpecComp}. 

\begin{figure*}
\resizebox{\hsize}{!}{\includegraphics[clip,angle=0]{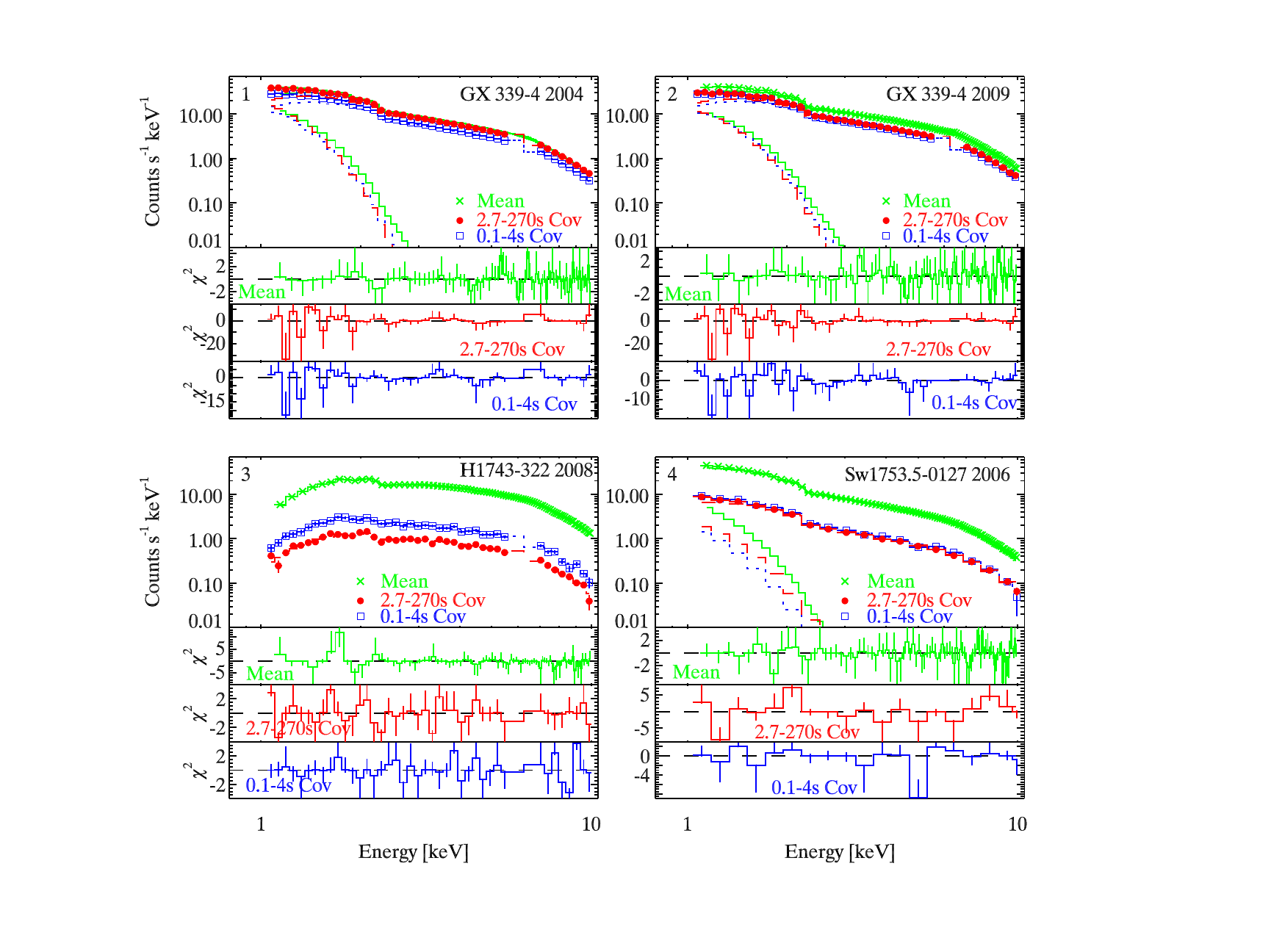}}
\caption{Mean energy spectra (green crosses) and covariance spectra on short (blue squares) and long (filled red circles) time scales for all eight observations. Best-fit model and individual components (disc blackbody and Comptonisation component) are indicated (dotted lines for short time scale covariance spectra; dashed lines for long time scale covariance spectra). In addition, $\chi^2$ distributions for the fits of all three spectra are shown.}
\label{Fig:cove_spec}
\end{figure*}

\addtocounter{figure}{-1}
\begin{figure*}
\resizebox{\hsize}{!}{\includegraphics[clip,angle=0]{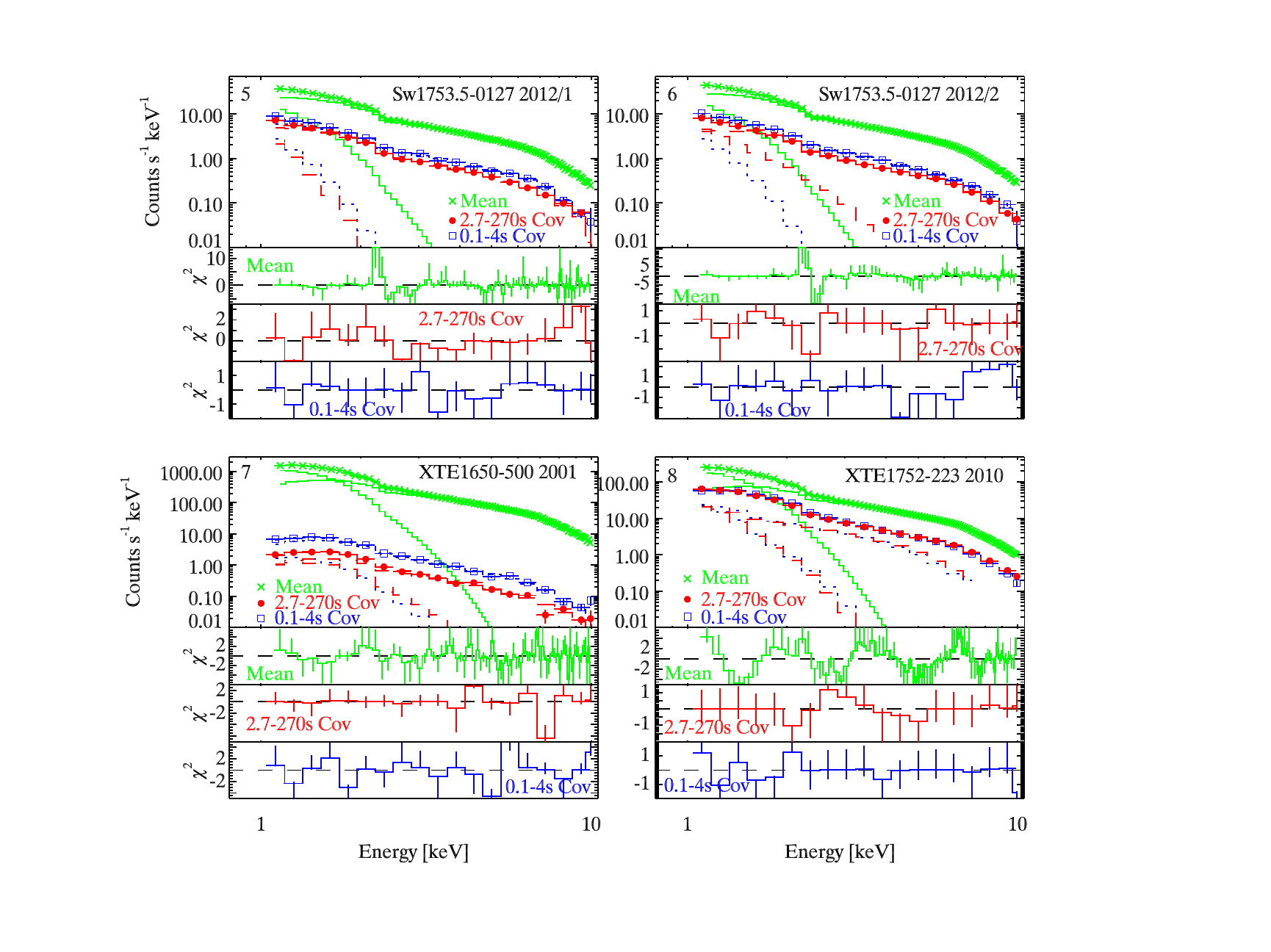}}
\caption{continued}
\end{figure*}

\begin{figure*}
\resizebox{\hsize}{!}{\includegraphics[clip,angle=0]{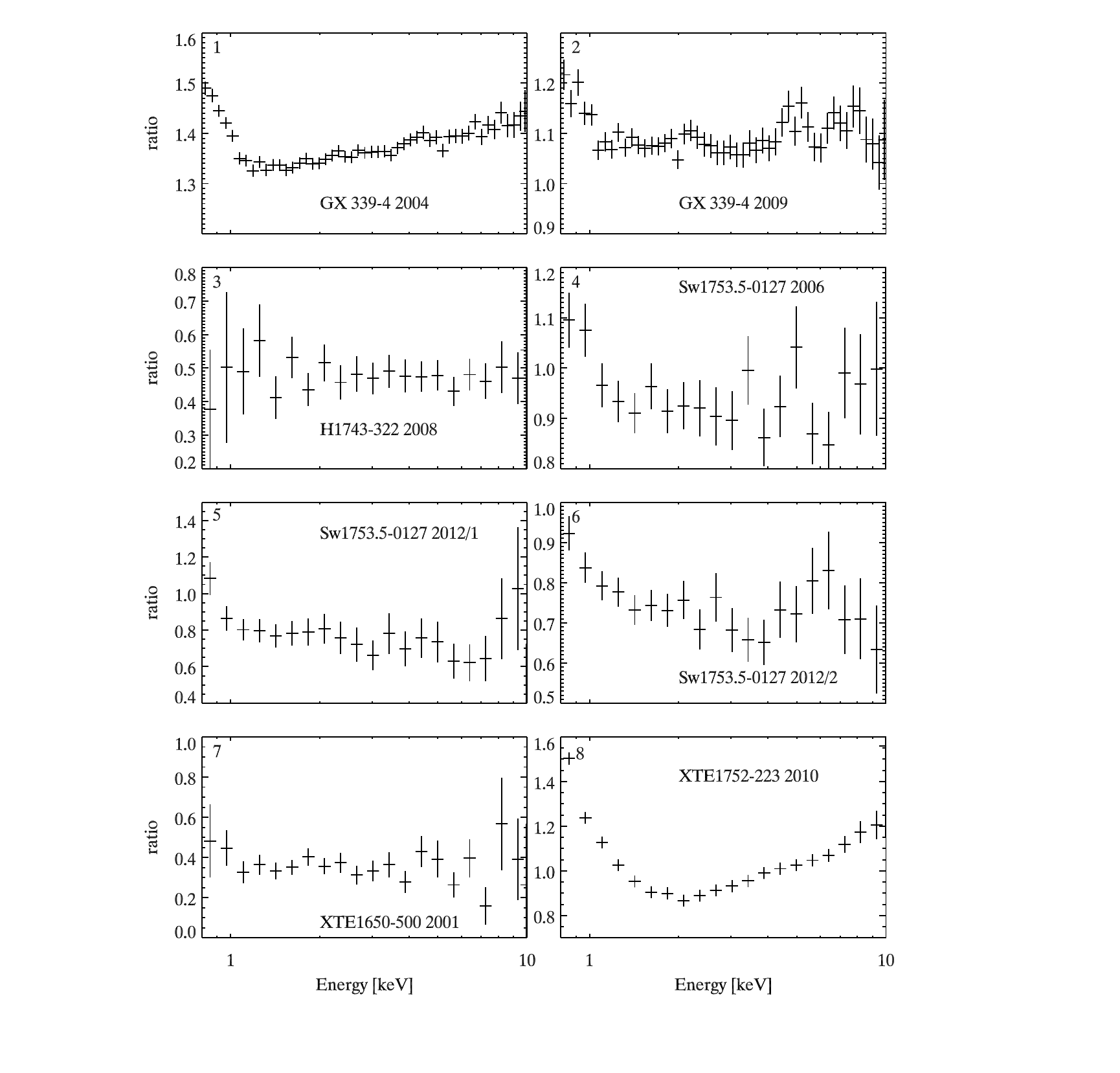}}
\caption{Covariance ratio of all eight observations.The ratios are obtained by dividing the covariance spectra obtained on longer time scales by those obtained on shorter time scales.}
\label{Fig:cov_ratio}
\end{figure*}

\begin{table*}
\caption{Parameters of covariance spectra}
\begin{center}
\begin{tabular}{lllllllll}
\hline\noalign{\smallskip}
 \multicolumn{1}{c}{param.} & \multicolumn{1}{c}{GX\,339/04} & \multicolumn{1}{c}{GX\,339/09}  & \multicolumn{1}{c}{Sw1753/06} &  \multicolumn{1}{c}{Sw1753/12/1}  &  \multicolumn{1}{c}{Sw1753/12/2}  &  \multicolumn{1}{c}{XTE1650}  &  \multicolumn{1}{c}{XTE1752} &  \multicolumn{1}{c}{H\,1743} \\
 \hline\noalign{\smallskip}
\multicolumn{9}{c}{short} \\
\hline\noalign{\smallskip}
N$_{\mr{dbb}}$ & $110485_{-31347}^{+27382}$ & $17357_{-4697}^{+8237}$  & $653_{-569}^{+104357}$ & $12876_{-11454}^{+459848}$ & $9763_{-7286}^{+96716}$ & $152_{-62}^{+275}$ & $846_{-121}^{+145}$ & -- \\
\noalign{\smallskip}
T$_{\mr{in}}$ [keV]& $0.183_{-0.002}^{+0.007}$ & $0.205\pm0.010$ & $0.198_{-0.086}^{+0.092}$ & $0.173_{-0.049}^{+0.040}$ & $0.179_{-0.046}^{+0.043}$ & $0.314_{-0.091}^{+0.058}$ & $0.382\pm0.012$ &  0.001$^{\natural}$\\
\noalign{\smallskip}
E$_{\mr{coff}}$ [keV]& $7.2\pm0.2$ & $>0.01$ & $>6.0$ & $7.5_{-1.2}^{+1.4}$ & $>5$ & $-$ & $-$ & $1.69_{-0.20}^{+0.14}$\\
\noalign{\smallskip}
E$_{\mr{fold}}$ [keV]& $19.7_{-1.1}^{+4.1}$ & $>0.01$ & $>0.01$ & $5.6_{-3.7}^{+16.6}$ & $>0.01$ & $-$ & $-$ & $2.5_{-0.2}^{+0.3}$\\
\noalign{\smallskip}
N$_{\mr{nthc}}$ & $0.324_{-0.005}^{+37.254}$ & $0.117_{-0.005}^{+0.004}$ & $0.013_{-0.002}^{+0.001}$ & $0.036_{-0.001}^{+0.003}$ & $0.048_{-0.006}^{+0.004}$ & $0.021_{-0.004}^{+0.007}$ & $0.071_{-0.003}^{+0.005}$ & $0.010\pm0.002$\\
\noalign{\smallskip}
$\Gamma$ & $1.58_{-0.03}^{+0.02}$ & $1.52\pm0.03$ & $1.68_{-0.08}^{+0.03}$ & $1.68_{-0.10}^{+0.05}$ & $1.75_{-0.05}^{+0.03}$ & $2.37_{-0.19}^{+0.09}$ & $2.08_{-0.04}^{+0.03}$ & $1.00_{-0.00}^{+0.46}$\\
\noalign{\smallskip}
kT$_{\mr{e}}$ [keV]& $7.1_{-1.5}^{+0.0}$ & $8.3_{-1.8}^{+3.7}$ & $>5$ & $>3.4$ & $>5.6$ & $>5$ & $>5$ & $<21$\\
\noalign{\smallskip}
$\chi^2$/dof& $109.7/31$ & $111.6/33$ & $22.8/12$ & $7.4/12$ & $14.9/12$ & $28.0/14$ & $4.4/8$ & $32.5/33$\\
 \hline\noalign{\smallskip}
\multicolumn{9}{c}{long} \\
\hline\noalign{\smallskip}
N$_{\mr{dbb}}$ & $272909_{-62781}^{+54522}$ & $29555_{-7352}^{+6530}$  & $6649_{-5371}^{+16426}$ & $28577_{-24309}^{+370948}$ & $6576_{-3258}^{+5856}$ & $24_{-9}^{+18}$ & $1124_{-168}^{+208}$ & -- \\
\noalign{\smallskip}
T$_{\mr{in}}$ [keV]& $0.169_{-0.005}^{+0.007}$ & $0.192_{-0.005}^{+0.008}$ & $0.154_{-0.018}^{+0.038}$ & $0.151_{-0.035}^{+0.000}$ & $0.199_{-0.017}^{+0.024}$ & $0.420_{-0.075}^{+0.065}$ & $0.361\pm0.011$ &  0.001$^{\natural}$\\
\noalign{\smallskip}
E$_{\mr{coff}}$ [keV]& $7.3_{-0.2}^{+0.3}$ & $>0.01$ & $>12.0$ & $>9.8$ & $>0.01$ & $-$ & $-$ & $<1.6$\\
\noalign{\smallskip}
E$_{\mr{fold}}$ [keV]& $22.1_{-3.1}^{+5.0}$ & $>0.01$ & $>0.01$ & $>0.01$ & $>0.01$ & $-$ & $-$ & $2.4_{-0.2}^{+1.0}$\\
\noalign{\smallskip}
N$_{\mr{nthc}}$ & $0.442_{-0.018}^{+0.011}$ & $0.133\pm0.002$ & $0.011\pm0.001$ & $0.034_{-0.000}^{+0.001}$ & $0.021_{-0.005}^{+0.002}$ & $0.005\pm0.002$ & $0.058_{-0.002}^{+0.003}$ & $0.004_{-0.001}^{+37.253}$\\
\noalign{\smallskip}
$\Gamma$ & $1.57_{-0.03}^{+0.02}$ & $1.54_{-0.03}^{+0.01}$ & $1.61_{-0.06}^{+0.03}$ & $1.77_{-0.04}^{+0.03}$ & $1.57_{-0.27}^{+0.05}$ & $2.24_{-0.29}^{+0.23}$ & $1.86_{-0.02}^{+0.03}$ & $1.00_{-0.00}^{+0.23}$\\
\noalign{\smallskip}
kT$_{\mr{e}}$ [keV]& $8.2_{-2.3}^{+4.3}$ & $7.0_{-1.3}^{+4.3}$ & $9.2_{-3.3}^{+72.2}$ & $>5.6$ & $>2.9$ & $>5$ & $>6.6$ & $<22$\\
\noalign{\smallskip}
$\chi^2$/dof& $143.5/31$ & $150.9/33$ & $23.9/9$ & $14.1/12$ & $8.5/12$ & $14.7/14$ & $5.8/8$ & $36.8/33$\\
\hline\noalign{\smallskip} 
\end{tabular} 
\end{center}
\label{Tab:CovSpecPar}
Notes: \\
$^{\natural}$: value used as seed temperature in the \textsc{nthcomp} model (fixed) 
\end{table*}

\begin{table*}
\caption{Behaviour of photon index and inner disc temperature comparing parameters obtained from covariance spectra on both time scales and of parameters obtained form  covariance spectra to those of mean energy spectra}
\begin{center}
\begin{tabular}{lllllll}
\hline\noalign{\smallskip}
 \multicolumn{1}{c}{param.} & \multicolumn{1}{c}{GX\,339/04} & \multicolumn{1}{c}{GX\,339/09}  & \multicolumn{1}{c}{Sw1753/06} &  \multicolumn{1}{c}{Sw1753/12/1}  &  \multicolumn{1}{c}{Sw1753/12/2}   &  \multicolumn{1}{c}{H\,1743} \\
 \hline\noalign{\smallskip}
\multicolumn{7}{c}{covariance to mean energy spectra} \\
\hline\noalign{\smallskip}
$\Gamma$ & smaller & agree & agree (s)  & agree (s) & $\dagger$ & agree\\
                  &              &           & smaller (l) & bigger (l) & &  \\
T$_{\mr{in}}$ & smaller & agree (s) & agree & smaller & smaller & smaller\\
                     &              &  smaller (l)                &  & & & \\
\hline\noalign{\smallskip}
\multicolumn{7}{c}{short term to long term covariance spectra} \\
\hline\noalign{\smallskip}
$\Gamma$ & agree & agree & agree & agree  & do not agree & agree \\
T$_{\mr{in}}$ & do not agree & agree  & agree  & agree  & agree  & agree \\
\hline\noalign{\smallskip} 
\end{tabular} 
\end{center}
\label{Tab:SpecComp}
Notes:\\
$^{\dagger}$: For the October 2012 observation of Swift\,J1753.5-0127 the photon index obtained from the mean energy spectrum lies between the values obtained from the covariance spectra and agrees within error bars with the one obtained from long scale covariance spectrum. \\
(s), (l): short, long term scale covariance spectra, respectively\\
agree: short for: agree within error bars\\
Table should be read like this: For the 2004 observation of \gx339\ the photon indices obtained from the long and short term covariance spectra are smaller than the one obtained from the mean energy spectrum.
\end{table*}

The consistency of photon indices of short and long time scale covariance spectra of the 2004 observation of \gx339\ and the 2006 observation of Swift\,J1753.5-0127 agrees with the results presented by \citet{2009MNRAS.397..666W}. They also obtained a bigger photon index from the mean energy spectrum compared to the one obtained from covariance spectra for the 2006 observation of Swift\,J1753.5-0127. In the case of the 2004 observation of \gx339\ \citet{2009MNRAS.397..666W} found a smaller photon index from the mean energy spectrum compared to the one obtained from covariance spectra, contrary to what we found. This difference can be related to the inclusion of a reflection component in the study of \citet{2009MNRAS.397..666W} that becomes obvious at higher energies covered by RXTE but inaccessible with \xmm. 

Apart form the observation of XTE\,J1650-500 and the October 2012 observation of Swift\,J1753.5-0127 the disc normalisation obtained from the long time scale covariance spectra are larger than the one obtained from the short time scale covariance spectra, although they all agree within their rather big error bars. In case of consistent photon indices, this higher disc blackbody normalisation indicates additional disc variability on long time scales \citep{2009MNRAS.397..666W}.

\subsubsection{Covariance ratios}
A model-independent way to compare variability on long and short time scales are covariance ratios, which we show in Fig.~\ref{Fig:cov_ratio}. The observations of \gx339, XTE\,J1752--223 and the 2006 and October 2012 observations of Swift\,J1753.5--0127 show a clear increase in the covariance ratio at energies below about one keV. Apart from the October 2012 observations of Swift\,J1753.5--0127 we found a higher disc blackbody normalisation for the long time scale covariance spectra in these observations, which is consistent with the increase found in the covariance ratios. This increase has been interpreted as a sign of additional disc variability on longer time scales by \citet{2009MNRAS.397..666W}. In case of the October 2012 observation of Swift\,J1753.5--0127 we found a change of the photon index between short and long time scale covariance spectra which may explain why the additional long-term disc variability observed in the covariance ratio does not show up in a higher disc normalisation in the long time scale covariance spectrum. We also found significantly different photon indices for the short and long time scale covariance spectra of XTE\,J1752--223. As changes in the photon index indicate changes of the overall spectral shape, the difference in photon index may explain the increase of covariance ratio with increasing energy seen in XTE\,J1752--223. 

\section[]{Discussion}
\begin{figure}
\resizebox{\hsize}{!}{\includegraphics[clip,angle=0]{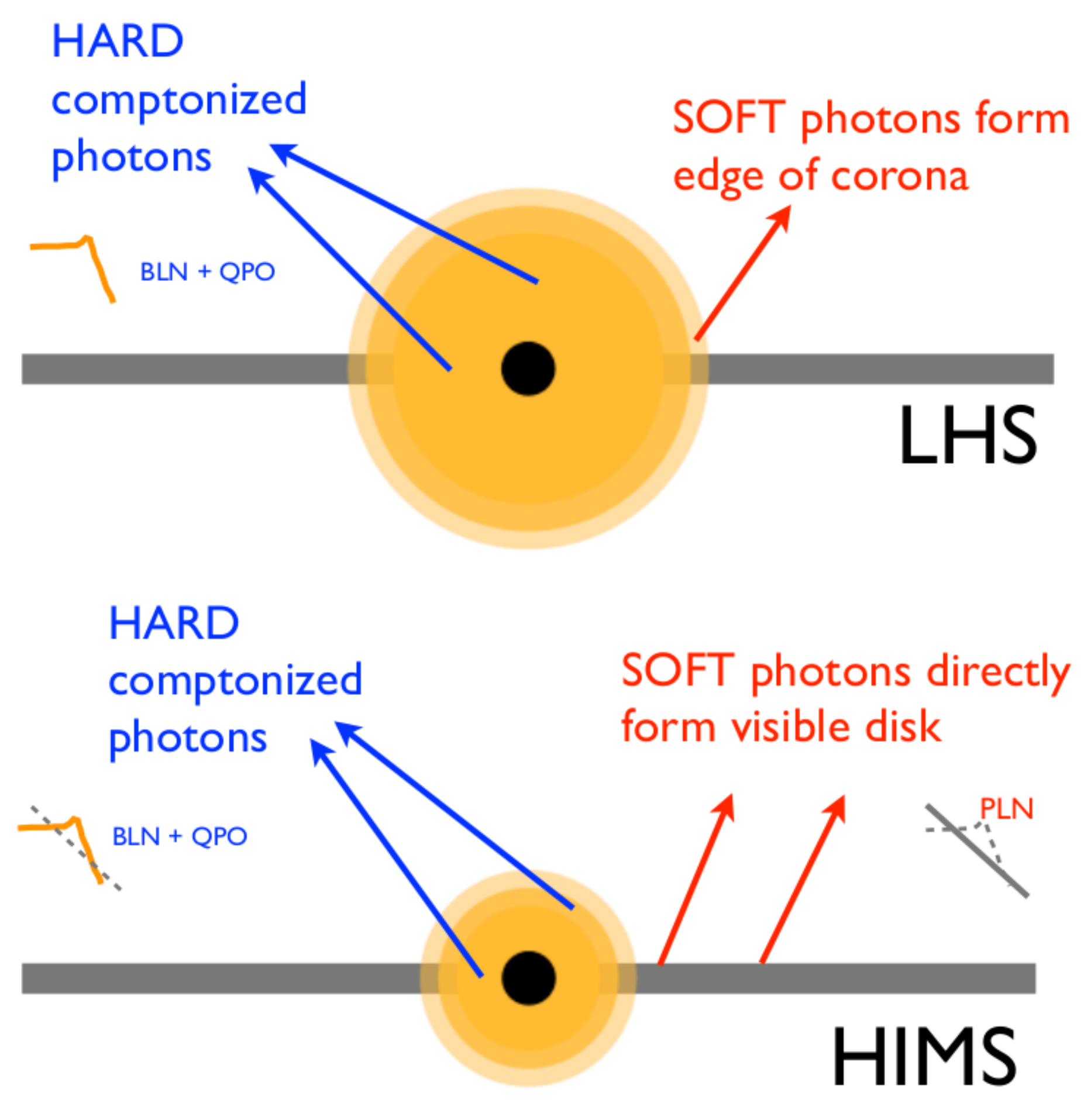}}
\caption{Schematic picture of the accretion geometry and the observed power spectral components obtained from our recent studies of black hole X-ray binaries during low hard state (upper panel) and around transition to the (hard) intermediate state (lower panel), see also \citet{2013ApJ...770..135Y} and \citet{2014MNRAS.441.1177S}.}
\label{Fig:sketch}
\end{figure}

At the beginning of this section we would like to summarise the main results:
\begin{itemize}
\item In Obs.\ 1, 2, 5, 6, 7, 8 the band limited noise component with the highest characteristic frequency shows a lower characteristic frequency in the soft band compared to the one observed in the hard band.
\item In all these observations the  ratio of the disc blackbody flux to the flux of the Comptonized component in the soft band exceeded $\sim$10\%. In Obs.\ 7 and 8 it even exceeds 100 per cent. In Obs.\ 3 and 4 the ratio is below 10 per cent.
\item In Obs.\ 1, 2, 3, 5 the rms spectrum is either flat or slowly decreasing with energy. Obs.\ 4 and 6 show a jump in variability around 4 -- 5 keV, while variability remains rather constant at lower and higher energies. Obs.\ 7 and 8 show much lower variability in the 1 -- 2 keV band compared to the rather constant variability above 2 keV.
\item The covariance ratios of Obs.\ 1, 2, 4, 6, 8 show a clear increase at lower energies. In Obs.\ 7 and 8 the photon indices and inner disc temperatures obtained from the covariance spectra are larger than those obtained from the mean energy spectra. Obs.\ 1, 2, 5, 6 do not show a general correlation between spectral parameters obtained from covariance and mean energy spectra.
\end{itemize}

Our investigations of the characteristic frequency of the band limited noise in a hard and soft energy band revealed that for observations where the ratio of the disc blackbody flux to the flux of the Comptonized component in the soft band exceeded $\sim$10\%, the band limited noise component with the highest characteristic frequency shows a lower characteristic frequency in the soft band compared to the one observed in the hard band. The energy dependence of the characteristic frequency is more obvious at characteristic frequencies above 1 Hz in the hard band.   

The energy dependence of the characteristic frequency observed in the 2004 observation of GX\,339--4 and the 2006 observation of Swift\,J1753.5-0127 concur with the energy dependence found in a qualitative study of the PDS shape that was based on the same \xmm\ observations \citep{2009MNRAS.397..666W}. This study of Wilkinson \& Uttley also presented the increase of covariance ratios at lower energies and interpreted it as additional variability of the accretion disc relative to the power-law component at low frequencies ($<$1 Hz), while the soft band blackbody variations at frequencies above 1 Hz are probably mostly produced by X-ray heating caused by the reprocessing of Comptonized photons in the accretion disc. 

In a study of XTE\,J1650-500 and XTE\,J1550-564, based on RXTE/PCA data, \citet{2005MNRAS.363.1349G} showed PDS in the $\sim$2 -- $\sim$13 and $\sim$13 -- $\sim$ 25 keV bands for different states. It is interesting to notice that for their LHS PDS of XTE\,J1650-500 the PDS of their hard band lies below the PDS of their soft band (\ie\ the power in the hard band is smaller than the power in the soft band at the same frequency), while for the energy bands used here the soft band PDS lies below the hard band PDS (see Fig.~\ref{Fig:PDS}).\@ In addition to the PDS  \citet{2005MNRAS.363.1349G} also showed rms spectra for different states. In their study of  XTE\,J1650-500 and XTE\,J1550-564 they observed flat rms spectra and an rms slightly decreasing with increasing energy in the LHS.\@ We found a similar behaviour in the observations of \gx339, \h1743, and the September 2012 observation of Swift\,J1753.5-0127. The behaviour of the rms observed in the remaining observations of Swift\,J1753.5-0127 corresponds to the one of the HIMS in the \citet{2005MNRAS.363.1349G} study.

The observed energy dependence of the characteristic frequency which is more obvious  at characteristic frequencies above 1 Hz in the hard band can be either explained by variable seed photon input for the Comptonized photons in different energy bands \citep{2005MNRAS.363.1349G} or by X-ray heating of the accretion disc  caused by the reprocessing of Comptonized photons \citep{2009MNRAS.397..666W}.
The variability properties observed in our study show characteristics of variability associated with Comptonized photons down to the lowest observed energies, \ie\ that the PDS can be described by BLN and QPOs in all energy bands and that a power-law shape observed in the HSS when disc variability dominates in not needed. The energy dependence of the characteristic frequency suggests that the seed photon input for the Comptonized photons varies between different energy bands, as has been discussed in \citet{2005MNRAS.363.1349G} based on RXTE data. It is therefore expected that such an effect would become more obvious when the corona is cooled down and the electron temperature is getting lower. Thus the energy dependence becomes more obvious in narrower energy bands.

The presence of the energy dependence of the characteristic frequency can be interpreted within the picture where the photons in the soft and hard energy bands come from different locations in the system. A schematic sketch of this picture, which was also used in our recent studies \citep{2013ApJ...770..135Y,2014MNRAS.441.1177S}, is given in Fig.~\ref{Fig:sketch}. When the thermal disc component emerges in the soft band of \xmm\ but does not dominate, a significant number of photons in the soft energy band will still come from the Comptonization component (originating either from an optically thin hot corona \citep[see \eg][]{1997ApJ...489..865E} or from a jet flow \citep[see \eg][]{2005ApJ...635.1203M}). However, the soft band photons should origin further away from the black hole than the hard band ones and thus suffer less up-scattering. This brings lower characteristic frequencies and softer energies together. The change of the characteristic frequency can be interpreted as a hint of a temperature gradient in the Comptonization component.

The observation that the energy dependence of the characteristic frequency is mainly observed at higher frequencies ($>$1 Hz) can be interpreted as a moving in of the inner disc radius during outburst evolution. In early outburst stages the disc ends far away from the black hole and the characteristic frequency, which is below 1 Hz, is consistent between both bands, since the photons in both bands experience a similar number of scatterings. With ongoing evolution of the outburst the boundary region between the disc and the Comptonizing corona moves inwards and the characteristic frequency as well as the contribution of the direct disc emission in the soft band increases \citep{2011MNRAS.415.2323I}. Notice that the radius at which the cold disc ends may be not the radius at which the cold disc terminates \citep[as assumed in the truncation disc model; \eg][]{2001A&A...373..251D,2006csxs.book..157M,2007A&ARv..15....1D}, since a hot flow or corona would cover the innermost cold disc so the radius one determines from the cut-off frequency of the power-law noise would correspond to the radius to which the hot flow or corona extends \citep[while the disc can reach down to the ISCO as assumed \eg\ in][]{1999ApJ...510L.123B,2002MNRAS.332..165M,2006ApJ...653..525M,2013ApJ...763...48R}. 

Interpreting the energy dependence of the characteristic frequency by additional intrinsic disc variability at low energies is not a convincing alternative, as intrinsic disc variability contributes at low frequencies \citep{2009MNRAS.397..666W} while the observed energy dependence of the characteristic frequency is more pronounced at frequencies above 1 Hz. In addition, the subsample of observations that show an increased covariance ratio at low energies does not agree with the subsample of observations that show the energy dependence of the characteristic frequency. We would like to point out that the additional disc variability on long time scales shows up only in the covariance spectra and ratios, but that there is no hint of an additional disc component in the PDS at low energies. As PDS fitting does not require any disc variability (see Sect.~\ref{SubSec:time_prop}), we can exclude that the observed energy dependence of the characteristic frequency is caused by underlying disc variability not taken into account properly during PDS fitting. Thus we prefer to interpret the energy dependence of the characteristic frequency to be caused by variable seed photon input for the Comptonized photons in different energy bands.

A challenge to our interpretation might be the observations of XTE J1650$-$500 and XTE J1752$-$223. In both observations the contribution of the disc emission to the overall emission in the soft band exceeds 50 per cent, and based on the results obtained in \citet{2014MNRAS.441.1177S} and \citet{2013ApJ...770..135Y} we would expect that the sources are in the hard intermediate state. The fact that the signal-to-noise ratio at higher energies is lower in these two observations compared to the other observations used in this study cannot explain why the contribution of the disc emission to the overall emission in the soft band exceeds 50 per cent, as taking the error bars on the spectral parameters into account does not affect the flux ratio to always exceed 50 per cent. We notice that for these two observations the variability amplitude in the softest band is lower than the variability amplitude at higher energies. Furthermore, fitting of the covariance spectra of these two sources does not require a high-energy cut-off and the photon indices and inner disc temperatures obtained from the covariance spectra are larger than those obtained from the mean energy spectra. For XTE J1752$-$223 this discrepancy can be explained by the fact that XTE J1752$-$223 is observed during outburst decay returning to the LHS, while the results of \citet{2014MNRAS.441.1177S} and \citet{2013ApJ...770..135Y} are based on the evolution from the hard to the soft state. In the case of XTE J1650$-$500, it is more difficult to find an explanation as the source is observed during outburst rise when it evolves from the LHS to the HSS. A possible explanation can be that in this source the contribution of the disc emission to the soft band is in general higher than in those sources studied by \citet{2014MNRAS.441.1177S} and \citet{2013ApJ...770..135Y}. A thorough investigation of this possibility as well as a detailed study of the differences between outburst rise and decay will be the targets of future studies. Please notice that in these two observations the signal-to-noise ratio at higher energies is lower compared to the other observations used in this study. However, the lower signal-to-noise ratio cannot explain why the contribution of the disc emission to the overall emission in the soft band exceeds 50 per cent.

Regarding the absence of correlations between the characteristic frequencies in different energy bands and between the characteristic frequency and spectral parameters, it is mainly the data points of \gx339\ with well constraint parameters that do not agree with possible trends seen in the remaining data points with bigger uncertainties. Higher quality data that would allow to constrain the characteristic frequency more precisely would be helpful to foster the absence of correlations.

In conclusion, the energy dependence of the break frequency of the band-limited noise component supports a picture that the power spectra in black hole X-ray binaries depend on which spectral components we are looking at. Energy-resolved timing observations can be sensitive enough to trace not only the accretion geometry by detecting power-law noise as evidence of the emergence of a disc component but can also probe the properties of the corona from studies of the band-limited noise break frequency. In this work we provide evidence that the energy dependence of the characteristic frequency may be used to probe the extension of the corona. 

\section*{Acknowledgments}
We would like to acknowledge useful discussions with Jon Miller, Mike Novak, Diego Altamirano, Omer Blaes and Tomaso Belloni. This work makes use of software tools provided by Simon Vaughan. This work was supported by the National Natural Science Foundation of China under grant No. 11073043, 11333005, and 11350110498, by Strategic Priority Research Program "The Emergence of Cosmological Structures" under Grant No. XDB09000000 and the XTP project under Grant No. XDA04060604, by the Shanghai Astronomical Observatory Key Project and by the Chinese Academy of Sciences Fellowship for Young International Scientists Grant. 

\bibliographystyle{mn2e}
\bibliography{}

%\appendix

%\section[]{Online Material}
%\include{}
\bsp

\label{lastpage}

\end{document}